\documentclass{iopart}
\newcommand{\EPL}{{\it Europhys. Lett.} }

\newcommand{\PRA}{{\it Phys. Rev.} A }

\newcommand{\OC}{{\it Opt. Commun.} }

\newcommand{\mycomm}[1]{}
\newcommand{\ashcomm}[1]{}

\newcommand{\UQ}{School of Mathematics and Physics, University of Queensland, Brisbane, 
Qld 4072, Australia.}
\usepackage{graphicx} 
\usepackage{epstopdf}                
\usepackage{bm}
\usepackage{iopams}
\usepackage{setstack}

\begin{document}

\title[Quantum CTAP and Entanglement]{Quantum dynamics and entanglement in coherent transport of atomic population}

\author{M.~K. Olsen}

\address{\UQ}

\begin{abstract} 
   
In this work we look at the quantum dynamics of the process known as either transport without transit (TWT), or coherent transfer of atomic population (CTAP), of a Bose-Einstein condensate from one well of a lattice potential to another, non-adjacent well, without macroscopic occupation of the well between the two. This process has previously been analysed and in this work we extend those analyses by considering the effects of quantum statistics on the dynamics and entanglement properties of the condensate modes in the two relevant wells. In order to do this, we go beyond the mean-field analysis of the Gross-Pitaevskii type approach and utilise the phase-space stochastic methods so well known in quantum optics. In particular, we use the exact positive-P representation where it is suitable, and the approximate truncated WIgner representation otherwise. We find strong agreement between the results of these two methods, with the mean-field dynamics not depending on the initial quantum states of the trapped condensate. We find that the entanglement properties do depend strongly on the initial quantum states, with quantitatively different results found for coherent and Fock states. Comparison of the two methods gives us confidence that the truncated Wigner representation delivers accurate results for this system and is thus a useful method as the collisional nonlinearity increases and the positive-P results fail to converge.

\end{abstract}


\submitto{\jpb}

\ead{mko@physics.uq.edu.au}

\maketitle

\section{Introduction}
\label{subsec:intro}

The possibility of macroscopic matter transport without transit (TWT) of a Bose-Einstein condensate (BEC) has been raised by Rab \etal~\cite{Rab2008,Bradly2012} and Opatrn\'{y} and Das~\cite{Opatrny2009}. The ideas behind this phenomenon come from the process of Stimulated Raman Adiabatic Passage (STIRAP), originally proposed to transfer atomic population between two atomic levels of a three-level atom without macroscopically populating an intermediate level~\cite{Oreg1984,Kuklinski1989,Bergmann1998}, and which involves the application of a counter-intuitive sequence of electromagnetic pulses to the atoms. Considering an atomic system with levels denoted by $|1\rangle,\: |2\rangle $ and $|3\rangle$, where the object is to transfer population from $|1\rangle$ to $|3\rangle$, a pulse denoted by $K_{23}$ dresses the $|2\rangle$ to $|3\rangle$ transition before a pulse denoted by $K_{12}$ is applied to the transition between $|1\rangle$ and $|2\rangle$. The result when everything works as proposed is that the population is transferred adiabatically from the first to the third state without populating the second.

Since the original proposal for the transfer of population between electronic levels, STIRAP has also been proposed for the conversion between atomic and molecular BEC, where there is one atomic state linked to an excited molecular state and finally a stable molecular state~\cite{Mackie2000,Hope2001,Drummond2002,Mackie2004,Ling2004}. A related Raman system with trapped BEC was proposed and analysed for a squeezed atom laser~\cite{Haine2005a,Haine2005b}, the production of atom-light entanglement~\cite{Haine2006}, quantum state measurement of an atom laser~\cite{Bradley2007}, and statistics swapping and quantum state transfer~\cite{Olsen2008}.

In this paper we will analyse the process of TWT in a three-well Bose-Hubbard system~\cite{Nemoto2000}, consisting of three potential wells in a linear configuration, where the condensed atoms are initially all in one of the end wells and at the final time have been transferred to the well at the other end, without significantly populating the middle well~\cite{Rab2008}. This system has previously been analysed using a Bose-Hubbard model~\cite{Rab2008,Bradly2012}, and a Gross-Pitaevskii model~\cite{Rab2008} which includes the spatial extent of the three wells with interactions between the atoms, as well as a Schr\"odinger equation approach with spatial extent but without interactions~\cite{Opatrny2009}. What we add here, by using a phase-space representation approach, is the quantum noise due to both the initial quantum statistics and the interactions, which allows us to calculate any deviations from the mean-field predictions~\cite{Chianca4well,Chiancathermal}, as well as the quantum correlations between the modes.

\section{System and Hamiltonian}
\label{sec:sysHam}

\begin{figure}[tbhp]
\begin{center} 
\includegraphics[width=0.9\columnwidth]{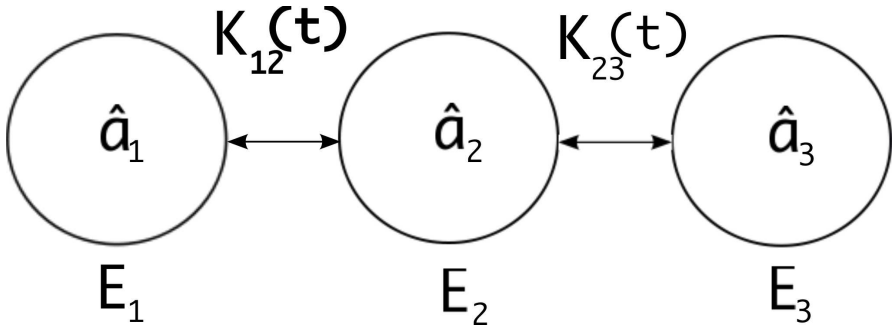}
\end{center}
\caption{Simpified schematic of the three well system. The $\hat{a}_{j}$ are bosonic annihilation operators for atoms in the $j$th well, the $E_{j}$ are the single-atom energy levels of each well, and the $K_{ij}(t)$ are the time dependent tunnelling strengths between wells $i$ and $j$.}
\label{fig:scheme}
\end{figure}

We may extend the procedure followed by Milburn \etal~\cite{Milburn1997} to treat our three-well system~\cite{Nemoto2000} as containing a single mode per well. A schematic is shown in Fig.~\ref{fig:scheme}, for which we can write the Hamiltonian,
\begin{eqnarray}
\eqalign{
{\cal H}/\hbar = \sum_{j=1}^{3}\left( E_{j}\hat{a}_{j}^{\dag}\hat{a}_{j} + \chi \hat{a}_{j}^{\dag\;2}\hat{a}_{j}^{2} \right)  \\
- K_{12}(t)\left(\hat{a}_{1}^{\dag}\hat{a}_{2}+\hat{a}_{1}\hat{a}_{2}^{\dag}\right) - K_{23}(t)\left(\hat{a}_{2}^{\dag}\hat{a}_{3}+\hat{a}_{2}\hat{a}_{3}^{\dag}\right),}
\label{eq:Ham}
\end{eqnarray}
where the $\hat{a}_{j}$ are the bosonic annihilation operators for atoms in each of the modes, the $E_{j}$ are the ground-state energies of each well, $\chi$ represents the collisional interactions, and the $K_{ij}(t)$ represent the time dependent couplings between the wells. Following Rab \etal~\cite{Rab2008}, we set $E_{1}=E_{3}=0$, $E_{2}>0$, with $K_{12}(t) = \Omega\sin^{2}\left[\pi t/2t_{p}\right]$ and $K_{23}(t) = \Omega\cos^{2}\left[\pi t/2t_{p}\right]$, where $t$ runs from $0$ to the pulse time, $t_{p}$. These parameters, with $\Omega = 10$, $t_{p}=400/\Omega$,  $E_{2}=\Omega$, and $-.005\leq \chi\leq 0.005$, were found to give good adiabaticity of population transfer, with very little population to be found in mode $2$ at any one time. We used a total number of atoms of $N_{A}=N_{1}+N_{2}+N_{3}=200$, where $N_{j}=\langle \hat{a}_{j}^{\dag}\hat{a}_{j}\rangle$. We note here that our Hamiltonian, although written in a slightly different way, is equivalent to that of Rab \etal~\cite{Rab2008}.

From the Hamiltonian we proceed in three different manners. The simplest is the Gross-Pitaevskii three-mode approach, which is semi-classical and, despite the fact that the process of condensation depends on quantum statistics, only allows us to calculate the mean field solutions. As has been shown many times, there is no guarantee that these are accurate for interacting atomic systems~\cite{JoeSuper,Chianca4well,Chiancathermal}. In many cases, the preferred option is then to use the positive-P representation~\cite{P+}, which allows for the exact calculation of normally ordered expectation values for systems described by this type of Hamiltonian, but can suffer from severe stability problems when the interaction term becomes dominant~\cite{Steel,unstableP+}. In cases where the instabilities prove insurmountable, a common option is to turn to the truncated Wigner representation~\cite{Graham1973}, which allows for the approximate calculation of symmetric expectation values. In this paper we will use and compare the results of all three of these methods. We note here that a recent development is a number-phase Wigner representation~\cite{Hush2010,Hush2012}, which is an improvement for the modelling of the interaction terms dependent on $\chi$, but has problems with the coupling terms. For this reason, we would not expect it to be an improvement over the truncated Wigner representation for our particular system. We will use the flexibility inherent in these phase space representations to model the initial state of the condensate in the first well as both a coherent state and a Fock state of fixed number~\cite{DFW}, in order to calculate the effects of the initial quantum statistics on both the subsequent dynamics and the quantum correlations of interest.

\section{Phase-space equations of motion and quantum state simulation}
\label{sec:EoM}

The semi-classical equations in the Gross-Pitaevskii approach are well known and can be found by several methods. They are written as
\begin{eqnarray}
\fl\eqalign{
\frac{d\alpha_{1}}{dt} = -i\left(E_{1}+2\chi|\alpha_{1}|^{2}\right)\alpha_{1} + i\Omega\sin^{2}\left(\pi t/2t_{p}\right)\alpha_{2}, \\
\frac{d\alpha_{2}}{dt} = -i\left(E_{2}+2\chi|\alpha_{2}|^{2}\right)\alpha_{2} + i\Omega\left[\sin^{2}\left(\pi t/2t_{p}\right)\alpha_{1} + \cos^{2}\left(\pi t/2t_{p}\right)\alpha_{3}\right] , \\
\frac{d\alpha_{3}}{dt} = -i\left(E_{3}+2\chi|\alpha_{3}|^{2}\right)\alpha_{3} + i\Omega\cos^{2}\left(\pi t/2t_{p}\right)\alpha_{2},}
\label{eq:BHGPE}
\end{eqnarray}
where we have included all the trap ground state energy levels, for generality. These semiclassical equations are useful for calculating the mean-field atom numbers in each well, $N_{j}=|\alpha_{j}|^{2}$, but can lose accuracy for all except short times~\cite{Chiancathermal}, and do not allow us to calculate any quantum correlations. To calculate these, we turn to the positive-P and truncated Wigner phase-space representations.

The truncated Wigner equations are found by truncating the third-order derivatives from the Wigner representation Fokker-Planck equation, which are due to the collisional terms in the Hamiltonian. This procedure leads to stochastic differential equations which are identical in form to the semiclassical set above. The difference, which allows us to calculate quantum correlations, is that the initial conditions are taken from the appropriate distribution for the quantum state we wish to consider~\cite{states}. We note that a method has been developed to model these third-order derivatives using stochastic difference equations, but that it is even more unstable than the positive-P representation~\cite{Wplus}. The truncated Wigner allows us to calculate approximately symmetrically ordered expectation values, such as, for example, 
\begin{equation}
\overline{|a_{j}|^{2}} \approx \frac{1}{2}\langle \hat{a}_{j}^{\dag}\hat{a}_{j}+\hat{a}_{j}\hat{a}_{j}^{\dag}\rangle = N_{j}+\frac{1}{2}.
\label{eq:Wignum}
\end{equation}
We note here that the line over the variables represents a classical average.

In order to develop the equations of motion for the positive-P phase-space variables, we proceed via the standard techniques~\cite{Crispin}, mapping the Hamiltonian onto a master equation, then the appropriate Fokker-Planck equations. Following the standard correspondence rules, we find the coupled stochastic differential equations in the positive-P representation, which necessitates six variables in a doubled phase-space to maintain positivity of the Fokker-Planck equation diffusion matrix,
\begin{eqnarray}
\fl\eqalign{
\frac{d\alpha_{1}}{dt} = -2i\chi\alpha_{1}^{2}\alpha_{1}^{+} + i\Omega\sin^{2}\left(\pi t/2t_{p}\right)\alpha_{2}+\sqrt{-2i\chi\alpha_{1}^{2}}\;\eta_{1}(t), \\
\frac{d\alpha_{1}^{+}}{dt} = 2i\chi\alpha_{1}^{+\;2}\alpha_{1} - i\Omega\sin^{2}\left(\pi t/2t_{p}\right)\alpha_{2}^{+} + \sqrt{2i\chi\alpha_{1}^{+\;2}}\;\eta_{2}(t), \\
\frac{d\alpha_{2}}{dt} = -i\left(E_{2}+2\chi\alpha_{2}^{+}\alpha_{2}\right)\alpha_{2} + i\Omega\left[\sin^{2}\left(\pi t/2t_{p}\right)\alpha_{1}+\cos^{2}\left(\pi t/2t_{p}\right)\alpha_{3}\right] +\sqrt{-2i\chi\alpha_{2}^{2}}\;\eta_{3}(t),\\
\frac{d\alpha_{2}^{+}}{dt} = i\left(E_{2}+2\chi\alpha_{2}^{+}\alpha_{2}\right)\alpha_{2}^{+} - i\Omega\left[\sin^{2}\left(\pi t/2t_{p}\right)\alpha_{1}^{+}+\cos^{2}\left(\pi t/2t_{p}\right)\alpha_{3}^{+}\right] +\sqrt{2i\chi\alpha_{2}^{+\;2}}\;\eta_{4}(t),\\
\frac{d\alpha_{3}}{dt} = -2i\chi\alpha_{3}^{2}\alpha_{3}^{+} + i\Omega\cos^{2}\left(\pi t/2t_{p}\right)\alpha_{2} + \sqrt{-2i\chi\alpha_{3}^{2}}\;\eta_{5}(t), \\
\frac{d\alpha_{3}^{+}}{dt} = 2i\chi\alpha_{3}^{+\;2}\alpha_{3} - i\Omega\cos^{2}\left(\pi t/2t_{p}\right)\alpha_{2}^{+} + \sqrt{2i\chi\alpha_{3}^{+\;2}}\;\eta_{6}(t).
}
\label{eq:P+SDE}
\end{eqnarray}
In the above, the $\eta_{j}$ are normal Gaussian noise terms, such that $\overline{\eta_{j}(t)}=0$ and $\overline{\eta_{j}(t)\eta_{k}(t')}=\delta_{jk}\delta(t-t').$ The averaging of the solutions of these equations over many stochastic trajectories approach normally-ordered operator expectation values, with, for example
\begin{equation}
\overline{\alpha_{j}^{+\;m}\alpha_{k}^{n}} \rightarrow \langle \hat{a}_{j}^{\dag\;m}\hat{a}_{k}^{n}\rangle ,
\end{equation}
whenever the integration is stable and converges. The positive-P representation is well known for instability and divergence properties, especially with $\chi^{(3)}$ nonlinearities~\cite{Steel,unstableP+}, but we will only be presenting results here where it has reliably converged.

The two initial quantum states which we will use are the Glauber-Sudarshan coherent state, which is the closest quantum state to a classical state of fixed amplitude and phase, and the Fock state, which has a fixed number but totally indeterminate phase~\cite{states}. We have chosen these two because, if the TWT is phase dependent, we would expect them to lead to the most different results. Both have been found previously to lead to results for three-well Bose-Hubbard systems which only match the classical predictions for short times~\cite{Chianca4well,Chiancathermal}, and have been shown to have  a marked effect on the dynamics of BEC molecular photoassociation~\cite{AMBEC2003,AMBEC2004,photoFock}. A coherent state, $|\alpha\rangle$, with coherent excitation $\alpha$, which includes the vacuum state, $|0\rangle$, is particularly easy to model in both representations. In the Wigner representation, we choose from the distribution
\begin{equation}
\alpha_{W} = \alpha +\frac{1}{2}\left(\eta_{1}+i\eta_{2}\right),
\label{eq:alphaWig}
\end{equation}
where the $\eta_{j}$ are sampled from a normal Gaussian distribution such that $\overline{\eta_{j}}=0$ and $\overline{\eta_{j}\eta_{k}}=\delta_{jk}$. We readily see that $\overline{\alpha_{W}}=\alpha$ and $\overline{|\alpha_{W}|^{2}}=|\alpha|^{2}+1/2$, as required. In the positive-P representation, one of the ways to sample $|\alpha\rangle$ is even simpler, with
\begin{equation}
\alpha_{P} = \alpha, \:\:\: \alpha_{P}^{+}=\alpha^{\ast}.
\label{eq:alphaP}
\end{equation}

A Fock state is a little more difficult, especially if we want an exact sampling in the Wigner representation, because it is not then drawn from a positive pseudo-probability distribution. However, as first noted by Gardiner \etal~\cite{Fudge}, it may be sampled to a close approximation for largish occupation numbers. A Fock state with $N$ quanta can be approximately modelled as ($N\gg 1$) by choosing the Wigner variables from the distribution
\begin{equation}
\alpha_{W} = (p+q\eta)\e^{2\pi i\nu},
\label{eq:Focksample}
\end{equation}
where
\begin{eqnarray}
\eqalign{
p = \frac{1}{2}\sqrt{2N+1+2\sqrt{N^{2}+N}}\\ 
q=\frac{1}{4p} ,}
\label{eq:pandq}
\end{eqnarray}
with $\eta$ being a standard Gaussian noise and $\nu$ being uniformly distributed on $[0,1]$. In the positive-P representation a Fock state can be sampled from
\begin{eqnarray}
\eqalign{
\alpha_{P} = \mu+\gamma, \\
\alpha_{P}^{+} = \mu^{\ast}-\gamma^{\ast},}
\label{eq:FockPplus}
\end{eqnarray}
where $\gamma=(\eta_{1}+i\eta_{2})/\sqrt{2}$, with the $\eta_{j}$ being normal Gaussian variables, and is easily sampled using the standard methods. The variable $\mu$ is found from a Gamma distribution for $z=|\mu|^{2}$,
\begin{equation}
\Gamma(z,N+1)=\frac{\mbox{e}^{-z}z^{N}}{N!},
\label{eq:Gammadist}
\end{equation}
using a method given by Marsaglia and Tsang~\cite{Marsaglia}, then taking $\mu=\sqrt{z}\;\mbox{e}^{i\theta}$, where $\theta$ is uniform on $[0,2\pi)$.

\section{Analysis of entanglement}
\label{sec:entangle}

In the present system we are interested in entanglement between the atomic modes at each site, rather than between individual atoms. In particular, we wish to investigate entanglement between the modes in wells $1$ and $3$, since these are the two in which the atomic populations are found. In practice this means that we are interested in whether the density matrix of the combination of modes can be written as a combination of density matrices for the individual modes, which follows the original idea as expounded by Schr\"odinger~\cite{Erwin,CVCpure}. For continuous-variable systems, the canonical methods use the system covariance matrix to develop inequalities based on the partial positive transpose criterion~\cite{Duan, Simon}, but we have previously found these to not be entirely suitable for systems with a $\chi^{(3)}$ nonlinearity, due to rotation of the Wigner function in phase-space~\cite{nonGauss}, which means that the quadrature angle for best measurement changes dynamically. 

We therefore turn to the work of Hillery and Zubairy~\cite{HZ}, which presents a wider range of inequalities, the violation of which can be used to demonstrate two-mode entanglement. Hillery and Zubairy used a Cauchy-Schwarz inequality to show that (using our notation),
\begin{equation}
\langle N_{1}N_{3}\rangle < |\langle \hat{a}_{1}\hat{a}_{3}^{\dag}\rangle |^{2},
\label{eq:HandZ}
\end{equation}
means that there exists bipartite entanglement between the atomic modes in wells $1$ and $3$. For our purposes, we define the function
\begin{equation}
\xi_{13} = \langle \hat{a}_{1}^{\dag}\hat{a}_{3}\rangle\langle \hat{a}_{3}^{\dag}\hat{a}_{1}\rangle -\langle\hat{a}_{1}^{\dag}\hat{a}_{1}\hat{a}_{3}^{\dag}\hat{a}_{3}\rangle,
\label{eq:xifunction}
\end{equation}
with the modes being entangled whenever $\xi_{13}>0$. Hilary and Zubairy show that this is a stronger measure than the Duan-Simon inequalities which are so useful for bipartite Gaussian entanglement, detecting entanglement in cases where it is missed by those measures. 

The calculation of $\xi_{13}$ is very straightforward using the positive-P representation, since it is already written in normal order. We find
\begin{eqnarray}
\eqalign{
\langle\hat{a}_{1}^{\dag}\hat{a}_{3}\rangle =\overline{\alpha_{1}^{+}\alpha_{3}}, \\
\langle\hat{a}_{3}^{\dag}\hat{a}_{1}\rangle = \overline{\alpha_{3}^{+}\alpha_{1}}, \\
\langle\hat{a}_{1}^{\dag}\hat{a}_{1}\hat{a}_{3}^{\dag}\hat{a}_{3}\rangle = \overline{\alpha_{1}^{+}\alpha_{1}\alpha_{3}^{+}\alpha_{3}}.
}
\label{eq:PPxi}
\end{eqnarray}
The equivalent calculation in the Wigner representation is slightly more complicated, with
\begin{eqnarray}
\eqalign{
\langle\hat{a}_{1}^{\dag}\hat{a}_{3}\rangle =\overline{\alpha_{1}^{\ast}\alpha_{3}}, \\
\langle\hat{a}_{3}^{\dag}\hat{a}_{1}\rangle = \overline{\alpha_{3}^{\ast}\alpha_{1}},\\
\langle \hat{a}_{1}^{\dag}\hat{a}_{1}\hat{a}_{3}^{\dag}\hat{a}_{3}\rangle = \overline{|\alpha_{1}|^{2}|\alpha_{3}|^{2}}-\frac{1}{2}\left(\overline{|\alpha_{1}|^{2}}+\overline{|\alpha_{3}|^{2}}\right)-\frac{1}{4}.
}
\label{eq:Wxi}
\end{eqnarray}
The use of the two representations will allow us to compare exact with approximate results and establish further the regime of validity of the truncated Wigner approximation for this system.

\section{Quantum dynamics}
\label{sec:dynamics}

\begin{figure}[tbhp]
\begin{center} 
\includegraphics[width=0.9\columnwidth]{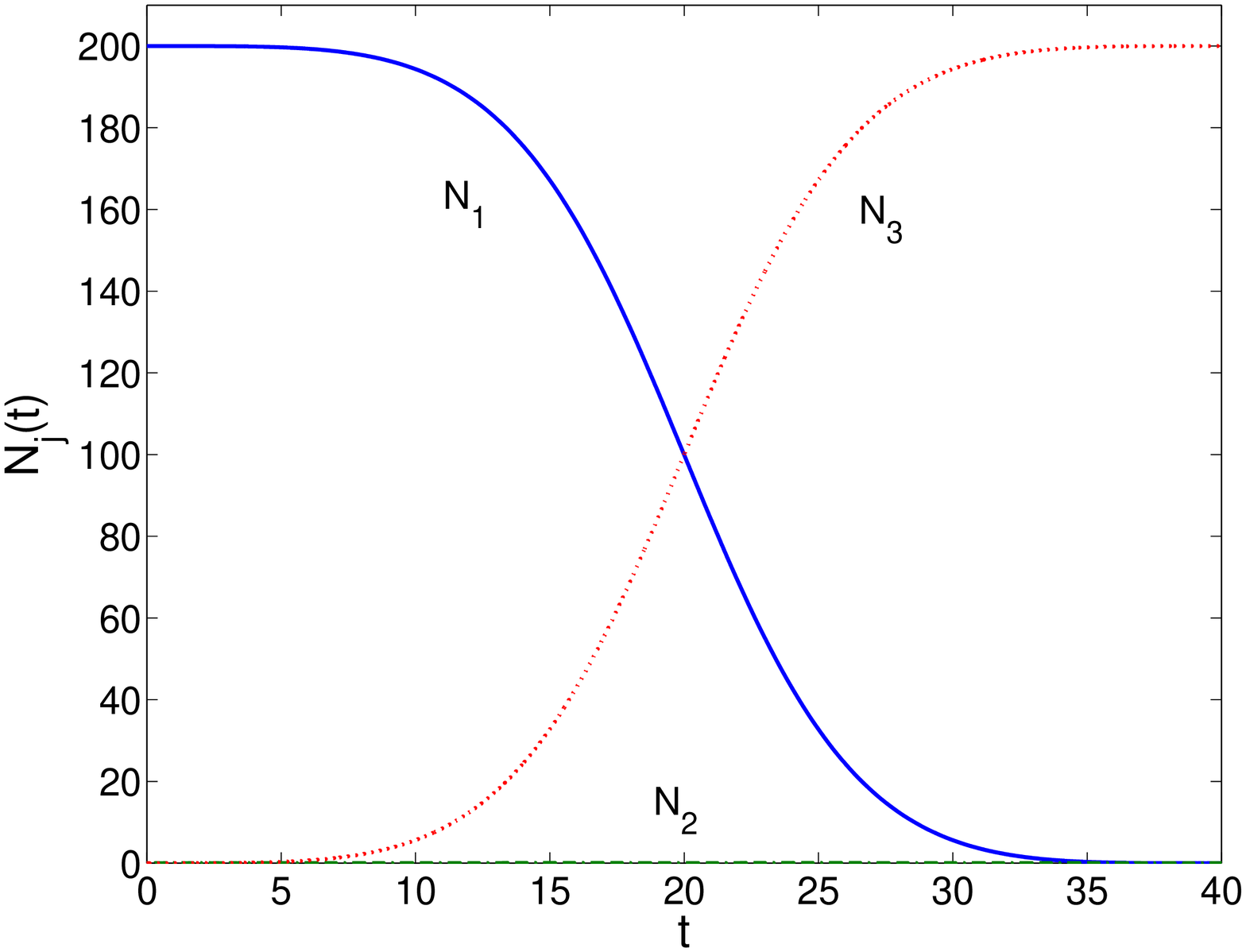}
\end{center} 
\caption{The populations in each of the three wells as a function of time (in units of $10/\Omega$). This graph was calculated using $6.51\times 10^{5}$ trajectories of the truncated Wigner equations, with $\Omega=10$, $E_{2}=1$, $\chi = 10^{-3}$, $N_{1}(0)=200$, $t_{p}=40$, and $N_{2}(0)=N_{3}(0)=0$. $N_{1}$ is initially in a coherent state. For these parameters, the GPE approach gives indistinguishable results for the populations, as does an initial Fock state in the first well.}
\label{fig:populations}
\end{figure}

In order to benchmark our results against what is already available, we have used parameters very similar to those of Rab \etal~\cite{Rab2008}. Our average number of atoms is $200$, with these all being found in the first well at $t=0$, and we have used $\Omega = 10$, with $t_{p}=400/\Omega$, and $E_{2}=0.1\Omega$ in all our calculations. Setting $E_{2}>0$ means that the coupling rate out of well $2$ into the adjacent wells is higher than that into it, thus helping to keep the well population as low as possible. We change $\chi$ and, in our stochastic simulations, the quantum state of the atomic mode in the first well. We find that the Gross-Pitaevskii approach gives good results, with a smooth one way transfer of population, and minimal population of the middle well, over the approximate range $-10^{-3}\leq \chi \leq 10^{-3}$, with the results for atomic number agreeing well with those calculated in the truncated Wigner representation over the full range. The truncated Wigner results for $\chi=10^{-3}$ are shown in Fig.~\ref{fig:populations}. The positive-P representation is not stable over this whole range, giving convergent results over $-.5\times 10^{-3}\leq \chi \leq .5\times 10^{-3}$. The initial quantum state in the first well has no discernible effect on the dynamics. These features, with the accuracy of the Gross-Pitaevskii approach and the independence of quantum state, are different to what we have found previously in two, three and four-well Bose-Hubbard models with normal couplings, and are most likely due to the shorter evolution time of the current model~\cite{Chianca4well,Chiancathermal}. We note here that the numbers of trajectories over which we integrate the stochastic equations are not the same for each set of parameters. This is for two reasons. The first is that the Fock state results converge with fewer trajectories than the coherent state results, which gives some indication that the natural state of the atoms in the first well is close to a Fock state. The second reason is in the way we write our programs for the stochastic integration, where the number of trajectories is not an input, but depends on how long we leave the programs running.

We next examine the entanglement properties of the system, concentrating on entanglement of the modes in the first and third wells. Entanglement with the middle well is not of interest here, as we require its population to stay as low as possible. It is obvious from inspection of Eq.~\ref{eq:xifunction} that it will have a value of zero when one of the modes involved has no atoms in it, or when both modes involved are in coherent states. We therefore expect to find any entanglement to be present during the middle period of the evolution, when the two outside wells are both populated. We note here that, due to the presence of the $\chi^{(3)}$ collisional nonlinearity, any entanglement present falls into the category of continuous-variable non-Gaussian entanglement, which has advantages over Gaussian entanglement for some quantum information tasks~\cite{nonGauss,Ohliger}.

\begin{figure}[tbhp]
\begin{center} 
\includegraphics[width=0.9\columnwidth]{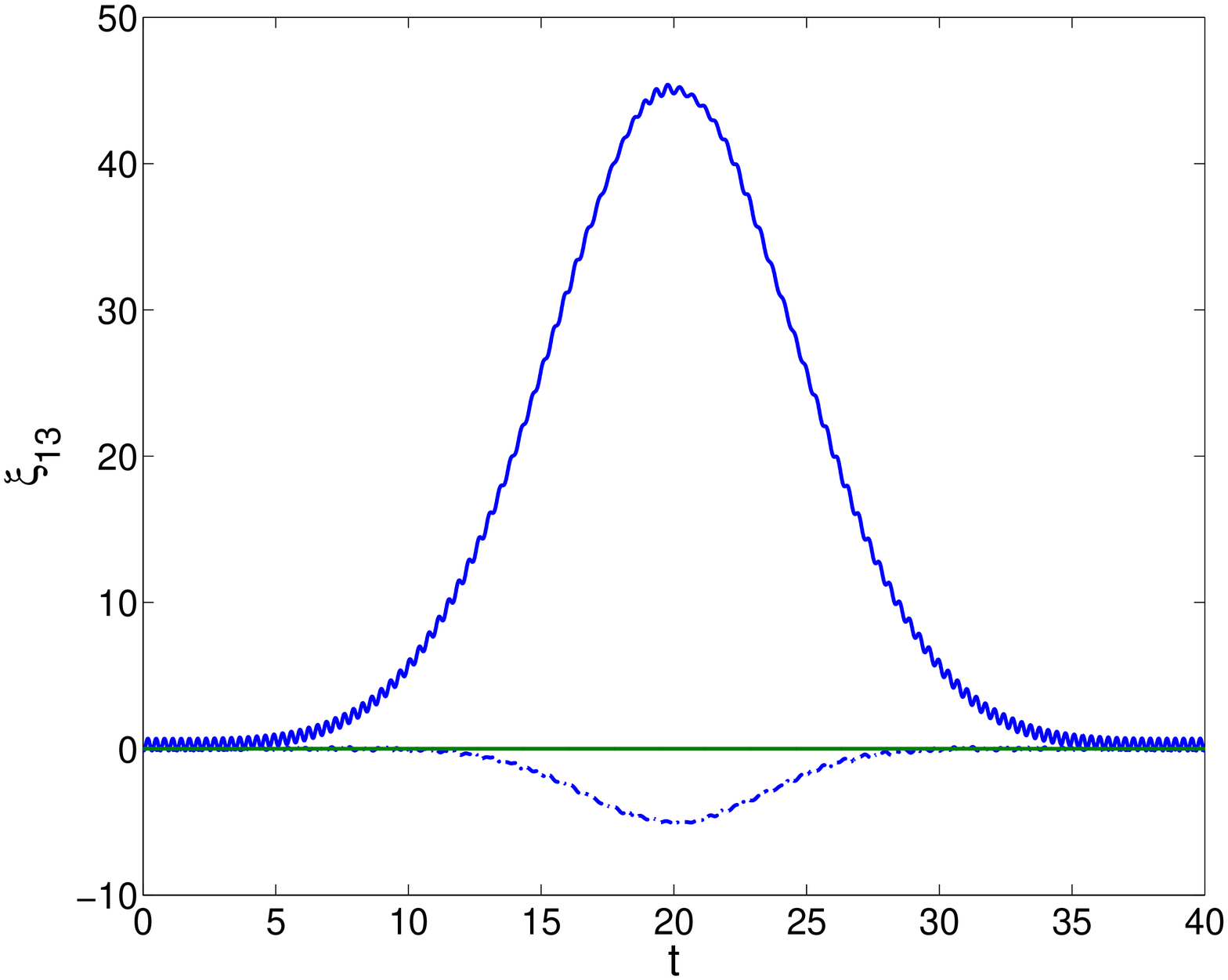}
\end{center} 
\caption{The entanglement correlation, $\xi_{13}$ of Eq.~\ref{eq:xifunction}, calculated for initial Fock (solid line) and coherent (dash-dotted line) states in the truncated Wigner representation. Parameters and initial conditions are as for Fig.~\ref{fig:populations}. The coherent state result is for $6.51\times 10^{5}$ trajectories and the Fock state result is for $2.66\times 10^{5}$. The solid line at $\xi_{13}=0$ is a guide to the eye.}
\label{fig:Wigxi}
\end{figure}

We expect that, to a first approximation, any entanglement will be maximised half way through the TWT process, when the numbers in the first and third wells are equal. If we assume that the quantum states remain the same through the evolution, i.e. the populations in these wells remain in either Fock or coherent states, with the occupation numbers changing, we can find analytic solutions for $\xi_{13}$ at this half way point. Although we do not expect these to be exact, they do give a benchmark and also an indication of how much the actual quantum states deviate from those at the beginning. For coherent states in each well, we find that $\xi_{13}=0$, which any deviation from this meaning that we no longer have two coherent states. For two Fock states with $N_{A}/2$ atoms in each well, we find $\xi_{13}=-N_{A}^{2}/4$, which also does not signify entanglement. This shows that any entanglement detected in this case means that the atoms have not remained in Fock states.

\begin{figure}[tbhp]
\begin{center} 
\includegraphics[width=0.9\columnwidth]{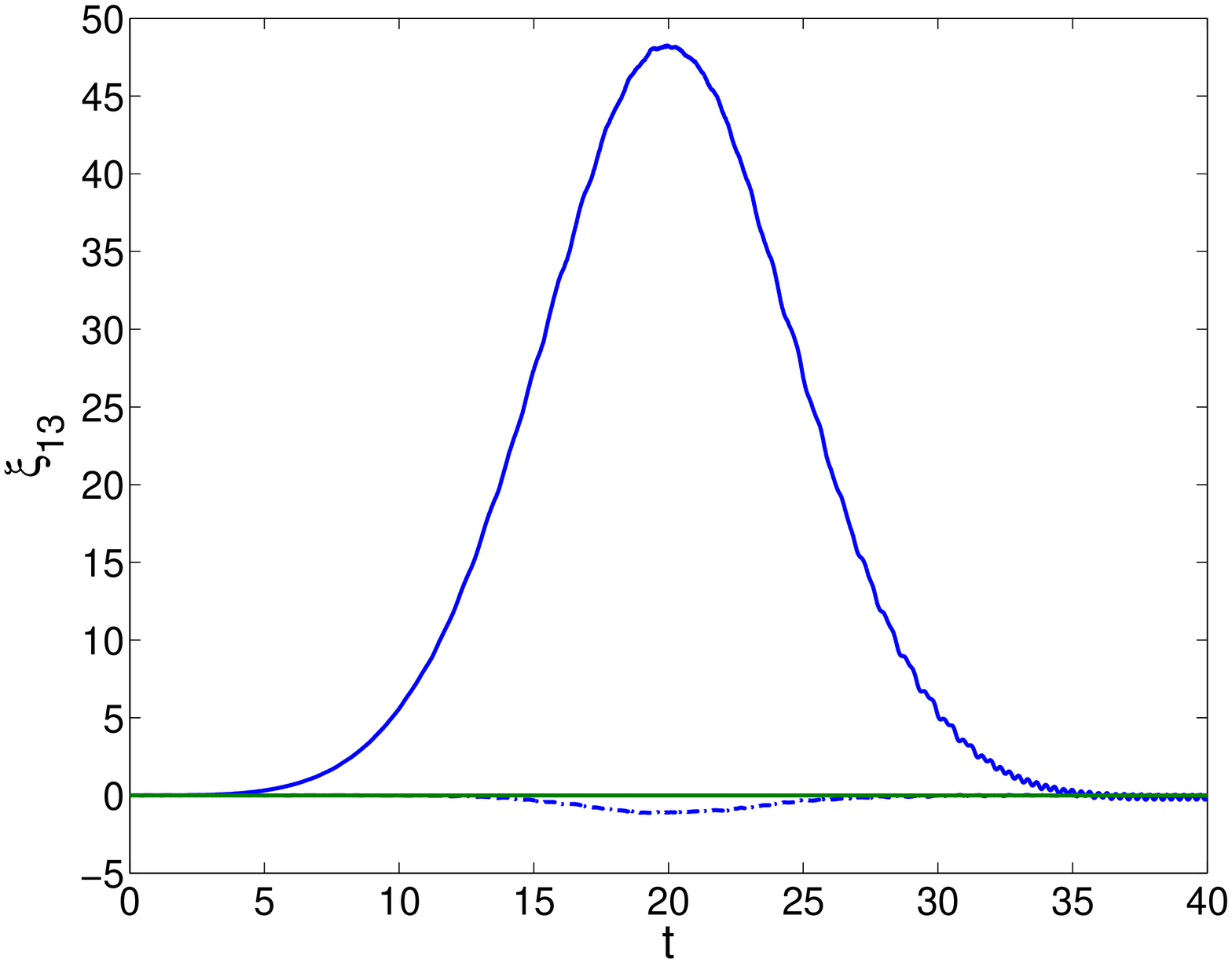}
\end{center} 
\caption{The entanglement correlation, $\xi_{13}$ of Eq.~\ref{eq:xifunction}, calculated for initial Fock (solid line) and coherent (dash-dotted line) states in the positive-P representation. The Fock state result is for $2.7\times 10^{4}$ trajectories, and the coherent state result is for $8.55\times 10^{5}$ trajectories. Parameters and initial conditions are as for Fig.~\ref{fig:populations}, but with $\chi=10^{-4}$. The solid line at $\xi_{13}=0$ is a guide to the eye.}
\label{fig:PPxi}
\end{figure}

In Fig.~\ref{fig:Wigxi} we show the results of truncated Wigner simulations for initial coherent and Fock states, both with $\chi=10^{-3}$, as in Fig.~\ref{fig:populations}. We immediately see that the absolute value of $\xi_{13}$ is a maximum for both cases at the half way point, where $t=t_{p}/2$, and is symmetric about this point. We also see that this measure does not demonstrate entanglement for an initial coherent state, but does for an initial Fock state. Fig.~\ref{fig:PPxi} shows the positive-P representation results, for $\chi=10^{-4}$ and everything else unchanged. We see that, except for the small amplitude oscillations in the truncated Wigner Fock state result, they are all very similar. These oscillations are presumably artefacts of the approximate nature of both the truncated WIgner equations and the representation of Fock states. Overall, the similarities give us confidence that the truncated Wigner representation is perfectly adequate for a quantum dynamical calculation of this system.

\section{Conclusions and discussion}
\label{sec:conclude}

In conclusion, we have performed both fully quantum and approximate analyses of the process of transport without transit for a three-well Bose-Hubbard type model where all the population is initially in one well. We find that the quantum statistics do not noticeably affect the mean-field dynamics for the representative cases of initial coherent and Fock states. We show that they do become important when we consider the entanglement properties of the system, with the dynamics following from an initial coherent state remaining separable, while an initial Fock state leads to entanglement of the atomic modes in the two end wells. 

The fact that the truncated Wigner representation gives results which are virtually indistinguishable from the exact positive-P representation in parameter regimes where the latter converges suggests that it is a perfectly adequate method for the analysis of this system. As well as presenting interesting quantum dynamics, this system is a source of non-Gaussian entangled states which are physically separated and composed of massive constituents. This could well make it useful for fundamental investigations of quantum information science with massive particles.

\ack

This research was supported by the Australian Research Council under the Future Fellowships Program (Grant FT100100515).

 

\Bibliography{99}

\bibitem{Rab2008}{Rab M, Cole J H, Parker N G, Greentree A D, Hollenberg L C L and Martin A M, 2008 \PRA {\bf 77} 061602(R)}
%
\bibitem{Bradly2012}{Bradly C J, Rab M, Greentree A D and Martin A M, 2012 \PRA {\bf 85} 053609}
%
\bibitem{Opatrny2009}{Opatrn\'{y} T and Das K K, 2009 \PRA {\bf 79} 012113}
%
\bibitem{Oreg1984}{Oreg J, Hioe F T and Eberly J H, 1984 \PRA {\bf 29} 690}
%
\bibitem{Kuklinski1989}{Kuklinski J R, Gaubatz U, Hioe F T and Bergmann K, 1989 \PRA {\bf 40} 6741}
%
\bibitem{Bergmann1998}{Bergmann K, Theuer H and Shore B W, 1998 \RMP {\bf 70} 1003}
%
\bibitem{Mackie2000}{Mackie M, Kowalski R and Javanainen J, 2000 \PRL {\bf 84} 3803}
%
\bibitem{Hope2001}{Hope J J, Olsen M K and Plimak L I, 2001 \PRA {\bf 63}, 043603}
%
\bibitem{Drummond2002}{Drummond P D, Kheruntsyan K V, Heinzen D J and Wynar R H, 2002 \PRA {\bf 65} 063619}
%
\bibitem{Mackie2004}{Mackie M, Harkonen K, Collin A, Suominen K A and Javanainen J, 2004 \PRA {\bf 70} 013614}
%
\bibitem{Ling2004}{Ling H Y, Pu H and Seaman B, 2004 \PRL {\bf 93} 250403}
%
\bibitem{Haine2005a}{Haine S A and Hope J J, 2005 \PRA {\bf 72} 033601}
%
\bibitem{Haine2005b}{Haine S A and Hope J J, 2005 Laser Phys. Lett. {\bf 2} 597}
%
\bibitem{Haine2006}{Haine S A, Olsen M K and Hope J J, 2006 \PRL {\bf 96} 133601}
%
\bibitem{Bradley2007}{Bradley A S, Olsen M K, Haine S A and Hope J J, 2007 \PRA {\bf 76} 033603}
%
\bibitem{Olsen2008}{Olsen M K, Haine S A, Bradley A S and Hope J J, 2008 Eur. Phys. J. Special Topics {\bf 160} 331}
%
\bibitem{Nemoto2000}{Nemoto K, Holmes C A, Milburn G J and Munro W J 2000 \PRA {\bf 63} 103604}
%
\bibitem{Chianca4well}{Chianca C V and Olsen M K, 2011 \PRA {\bf 83} 043607}
%
\bibitem{Chiancathermal}{Chianca C V and Olsen M K, 2011 \PRA {\bf 84} 042636}
%
\bibitem{Milburn1997}{Milburn G J, Corney J F, Wright E M and Walls D F, 1997 \PRA {\bf 55} 4318.}
%
\bibitem{JoeSuper}{Hope J J and Olsen M K, 2001 \PRL {\bf 86} 3220}
%
\bibitem{P+}{Drummond P D and Gardiner C W 1980 \JPA {\bf 13} 2353}
%
\bibitem{Steel}{Steel M J, Olsen M K, Plimak L I, Drummond P D, Tan S M, Collett M J, Walls D F and Graham R, 1998 \PRA {\bf 58} 4824}
%
\bibitem{unstableP+}{Plimak L I, Olsen M K and Collett M J, 2001 \PRA {\bf 64} 025801}
%
\bibitem{Graham1973}{Graham R, 1973 Springer Tracts Mod. Phys. {\bf 66} 1}
%
\bibitem{Hush2010}{Hush M R, Carvalho A R R and Hope J J, 2010 \PRA {\bf 81} 033852}
%
\bibitem{Hush2012}{Hush M R, Carvalho A R R and Hope J J, 2012 \PRA {\bf 85} 023607}
%
\bibitem{DFW}{Walls D F and Milburn G J, \emph{Quantum Optics} (Springer-
Verlag, Berlin, 1995).}
%
\bibitem{states}{Olsen M K and Bradley A S, 2009 \OC {\bf 282} 3924}
%
\bibitem{Wplus}{Plimak L I, Olsen M K, Fleischhauer M and Collett M J, 2001 \EPL {\bf 56} 372}
%
\bibitem{Crispin}{Gardiner C W \emph{Quantum Noise} (Springer, Berlin,
1991)}
%
\bibitem{AMBEC2003}{Olsen M K and Plimak L I, 2003 \PRA {\bf 68} 031603}
%
\bibitem{AMBEC2004}{Olsen M K, 2004 \PRA {\bf 69} 013601}
%
\bibitem{photoFock}{Olsen M K, Bradley A S and Cavalcanti S B, 2004 \PRA {\bf 70} 033611}
%
\bibitem{Fudge}{Gardiner C W, Anglin J R and Fudge T I A, 2002 \JPB {\bf 35} 1555}
%
\bibitem{Marsaglia}{Marsaglia G and Tsang W W, 2000 ACM Trans. Math. Software {\bf 26} 363}
%
\bibitem{Erwin}{Schr\"odinger E, 1935 Proc. Cambridge Philos. Soc. {\bf 31} 555}
%
\bibitem{CVCpure}{Chianca C V and Olsen M K, 2012 \OC {\bf 285} 825}
%
\bibitem{Duan}{Duan L-M, Giedke G, Cirac J I and Zoller P, 2000 \PRL {\bf 84} 2722}
%
\bibitem{Simon}{Simon R, 2000 \PRL {\bf 84} 2726}
%
\bibitem{nonGauss}{Olsen M K and Corney J F, 2013 \PRA {\bf 87} 033839}
%
\bibitem{HZ}{Hillery M and Zubairy M S, 2006 \PRL {\bf 96} 050503}
%
\bibitem{Ohliger}{Ohliger M. Kieling K and Eisert J, 2010 \PRA {\bf 82} 042336}
%



\endbib
\end{document}